\documentclass[11pt]{article}
\usepackage[title]{appendix}
\usepackage{epsfig, amsmath, amssymb, amsthm, times, stackrel, bm}
\usepackage{color}
\usepackage{esint}

\usepackage{xcolor}
\usepackage{hyperref}

\newcommand{\field}[1]{\mathbb{#1}}

\newcommand{\N}{\field{N}}    % natural integers
\newcommand{\Z}{\field{Z}}    % integers
\newcommand{\R}{\field{R}}    % real numbers
\newcommand{\cW}{{\mathcal W}}  % calligraphic W
\newcommand{\bW}{\mathbf{W}}

\newcommand{\1}{\mathbf{1}}  
\def\be{\begin{equation}}
  \def\ee{\end{equation}}

\newtheorem{theorem}{Theorem}[section]

\newtheorem{remark}[theorem]{Remark}

\title{Graphon Signal Processing for Spiking and Biological \\Neural Networks
}
\date{}
\author{Takuma Sumi\thanks{Hotchkiss Brain Institute, Cumming School of Medicine, University of Calgary, Calgary, Alberta, Canada}  \and  Georgi S. Medvedev\thanks{
   Department of Mathematics, Drexel University, Philadelphia, PA 19002, USA} 
}

\begin{document}
\maketitle
\begin{abstract}
 Graph Signal Processing (GSP) extends classical signal processing to
signals defined on graphs, enabling filtering, spectral analysis,
and sampling of data generated by networks of various kinds.
Graphon Signal Processing
(GnSP) develops this framework further by employing the theory of
graphons. \textit{Graphons} are measurable functions on the unit square
that represent graphs and limits of convergent graph sequences.
The use of graphons provides stability of GSP methods to stochastic
variability in network data and improves computational efficiency
for very large networks.

We use GnSP to address \textit{the stimulus identification problem} (SIP) in
computational and biological neural networks. The SIP is an inverse problem
that aims to infer the unknown stimulus \(s\) from the observed network
output \(f\). We first validate the approach
in spiking neural network simulations and then analyze calcium imaging
recordings. Graphon-based spectral
projections yield trial-invariant, low-dimensional
embeddings that improve stimulus classification over Principal Component
Analysis and discrete GSP baselines. The embeddings remain stable under
variations in network stochasticity,
providing robustness to different network sizes and noise levels.
To the best of our knowledge, this is the first application of GnSP
to biological neural networks, opening new avenues for graphon-based
analysis in neuroscience.
\\

\end{abstract}

\section{Introduction}
\label{sec:1}

Graph Signal Processing (GSP) is a powerful extension of classical signal processing
techniques oriented towards data defined on irregular graph domains. It  provides
tools for filtering,
spectral analysis, and sampling adapted to data generated by complex networks of
different nature
\cite{Leus2023,Ortega2018,Shuman2013}. This set of techniques provides
systematic ways to analyze network data, but its results in general
depend strongly on the given graph structure, making it challenging at times
to generalize across similar but not identical graphs.
More recently, Graphon Signal Processing (GnSP) has been proposed
to extend GSP to the continuous setting by using graphons, measurable
functions on the unit square which represent graphs and limit points of graph sequences
\cite{Ruiz2021, Lovasz-book}. The use of graphons helps to capture the principal spectral
properties
intrinsic to the given network and
supports size‐ and noise‐robust analysis for graphs of increasing size.
Despite these advantages, GnSP has largely been applied to
synthetic data and theoretical studies in network science,
with few studies addressing experimental data, where
trial-to-trial variability and finite-size effects pose
significant challenges. The present study aims to close this gap.

Advanced microfabrication \cite{MontalaFlaquer2022,Sumi2025}, microcontact printing
\cite{Vogt2003,Albers2016,Yamamoto2016,Yamamoto2018}, and microfluidic \cite{Millet2012,Bisio2014,Pan2015}
techniques enable the creation of cultured neuronal networks with precisely engineered
architectures. Neurons can be organized into modular layouts
\cite{Habibey2024,WinterHjelm2023,Yamamoto2018,Sumi2023},
with densely connected clusters linked by sparse inter‐cluster connections—a motif
conserved across brain networks \cite{Meunier2010,Sporns2016}. These well-controlled
\textit{in vitro} networks offer an ideal testbed for applying GnSP to extract
stable, low-dimensional representations and to assess the robustness of graphon-level
spectral features under biological variability.

The goal of this paper is to apply GnSP to address
the inverse problem of inferring stimulus patterns \(s\) from observed neural responses
\(f\) in biological neural networks. We validate the proposed method first in numerical simulations of
networks of integrate-and-fire neurons, and then apply it to calcium‐imaging data from
cultured modular networks.
Using three different stimulation patterns \(s_1\), \(s_2\), and \(s_3\), we show that GnSP
yields trial‐invariant, separable low‐dimensional maps and achieves higher classification accuracy
than standard dimensionality reduction methods, such as Principal Component Analysis (PCA)
and discrete GSP baselines. The embeddings remain stable under variability in network size and
stochasticity, enabling size‐ and noise‐robust analysis. To our knowledge, this is the first
application
of GnSP to experimental neuroscience, opening the door to its broader use in neural
data analysis.

The organization of the paper is as follows. In the next section, we provide the reader with
the necessary background on GSP and GnSP. In Section~\ref{sec.block}, we define the stochastic
block network used in the spiking neural network simulations. The structure of this model is
inspired
by the cultured neuronal networks designed and analyzed in \cite{Sumi2023}. In
Section~\ref{sec.spiking},
we introduce the spiking neural network and present numerical simulations of its dynamics.
In Section~\ref{sec.sip}, we apply GnSP to identify stimuli in numerical simulations of the
spiking network model under two protocols: a two‐stimulus protocol, where different clusters
are selectively stimulated, and a mixed‐stimulus protocol, where multiple clusters can be
partially stimulated. For both cases, the task is to identify the stimulated regions from
the network’s output. In Section~\ref{sec.experiment}, we apply the same framework to
experimental calcium‐imaging data from \cite{Sumi2023}. The results on experimental data
are consistent with the simulations: in both settings, GnSP provides a simple and efficient
method for stimulus identification. 
In Section~\ref{sec.generalization}, we extend the GnSP framework beyond the four-block model to small-world and higher-block graphons to demonstrate its robustness and scalability.
Finally, in Section~\ref{sec.discuss} we summarize our
results and discuss broader implications for graph‐based methods in neuroscience.

\section{Graphs, graphons, and signal processing}\label{GSP}
\setcounter{equation}{0}

In the main part of this paper, we apply GnSP to address the stimulus
identification problem (SIP) in computational and biological neural networks.
To set the stage, this section
reviews some basic facts about GnSP, which will be used in the remainder of this paper.

Let $\Gamma^n=\langle V(\Gamma^n), E(\Gamma^n)\rangle$ be a sequence of undirected
simple connected graphs. $V(\Gamma^n)$ and $E(\Gamma^n)$ stand for the node and edge
sets of $\Gamma^n$
respectively. We label the nodes of $\Gamma^n$ by the natural numbers from $1$ to $a_n$:
$$
V(\Gamma^n)=[a_n]:=\{1,2,\dots, a_n\}, \quad a_n\nearrow \infty.
$$
The adjacency matrix of $\Gamma^n$ is denoted by $A^n$. It is an $a_n\times a_n$ symmetric
matrix $A^n=(A^n_{ij})$ (cf.~\cite{Biggs-book})
$$
A^n_{ij}=\left\{ \begin{array}{ll} 1, & i\sim j,\\
                   0, & \mbox{otherwise}.
                 \end{array}
               \right.
               $$
               We list the eigenvalues of $A^n$ in the descending order counting multiplicity:
               $$
               \lambda_1(A^n)\ge  \lambda_2(A^n)\ge\dots \ge \lambda_{a_n}(A^n).
               $$
               The corresponding eigenvectors (after normalization) form an orthonormal basis
               in $\R^{a_n}$:
               $$
                    v_1(A^n), v_2(A^n),\dots, v_{a_n}(A^n).
               $$
               
               Let $f$ be a function from $V(\Gamma^n)$ to the real line,
               $f\in L\left(V(\Gamma^n), \R\right)$. $f$ is interpreted as a signal. In the context
               of the SIP, $f\in\R^{a_n}$ codes the response of the network to a given stimulus.
               In GSP, one decomposes $f$ with respect to the eigenbasis of $A^n$:
              \begin{equation}\label{F-expansion}
                f=\sum_{i=1}^{a_n} \left( f,  v_i(A^n)\right) v_i(A^n).
                \end{equation}

Notably, the first few Fourier coefficients in \eqref{F-expansion} already
capture essential information, sufficient for solving many practical problems
\cite{Ortega-book}. This is the central idea of GSP. For the SIP, we will
discuss in detail the geometric meaning of these leading Fourier coefficients below.

For sequences of random graphs $\Gamma^n$ that commonly arise in applications,
different realizations of $\Gamma^n$ yield eigenspaces that are typically distinct,
yet remain close to one another. In fact, for the random graph model considered below,
the sequence $(\Gamma^n)$ is, with high probability, convergent in the sense of dense
graph convergence (cf.~\cite{Lovasz-book}). The eigenspaces of the corresponding
graph limit, a kernel operator on $L^2([0,1])$, approximate those of $A^n$ for
large $n$, in a sense that will be made precise below.

This allows us to use the eigenspaces of the limiting operator to compute Fourier coefficients
across different realizations of $A^n$, as well as for all sufficiently large $n$. Beyond this,
graph limits offer additional advantages for signal processing, which we will discuss later in the
context of the SIP. The extension of graph
signal processing that builds on graph limits is referred to as \textit{graphon signal processing} (GnSP) \cite{Ruiz2021, Morency2021}.

Before continuing, we briefly review a few basic ideas from the theory of
graph limits, sufficient for our purposes. For further details on this rich
and versatile theory, with its many applications, we refer the interested
reader to \cite{Lovasz-book}.

  In the graph limit theory, functions are used to represent graphs and the limit points of graph
  sequences. Such functions are called \textit{graphons}. The space of graphons equipped
  with \textit{cut-norm} metric provides a convenient setting for 
  studying convergence of graph sequences using analytical techniques. For the problem
  at hand, this scheme can be implemented as follows.

  First, we split $[0,1]$ into $a_n$ subintervals:
  \begin{equation}\label{def-I}
  I^n_1=\left[0, \frac{1}{a_n}\right), I^n_2=\left[\frac{1}{a_n}, \frac{2}{a_n}\right),\dots,
  I^n_i=\left[\frac{i-1}{a_n}, \frac{i}{a_n}\right), \dots,
  I^n_{a_n}=\left[\frac{a_n-1}{a_n}, 1\right].
  \end{equation}
  Then we introduce
  \begin{equation}\label{def-Wn}
  W^n(x,y)=\sum_{i,j=1}^{a_n} A^n_{ij}\1_{ I^n_i}(x) \1_{ I^n_j}(y),
  \end{equation}
     where $\1_B$ stands for the indicator function of set $B$.
     Function $W^n$ is a graphon representing
     $\Gamma^n$.

     Let $\cW_0$ be a set of bounded measurable symmetric functions on the unit square. This is the
     space of graphons. The cut-norm is used to measure the distance between functions from
     $\cW_0$. In practice, it is often more convenient to work with the $\infty\to 1$ norm, which is
     equivalent to the cut norm. The $\infty\to 1$ norm is defined as follows
    \begin{equation}\label{cut-norm}
\|U\|_{\infty\to 1}=\sup_{|f|,|g|\le 1} \left| \int_{[0,1]^1} f(x) U(x,y) g(y) dxdy \right|.
\end{equation}
The supremum in \eqref{cut-norm} is taken over all bounded measurable functions on $[0,1]$ not
exceeding $1$ in absolute value.

     We say that $(\Gamma^n)$ is a convergent sequence with limit $W\in\cW_0$ if
\begin{equation}\label{cut-conv}
\lim_{n\to\infty} \|W^n-W\|_{\infty\to 1} =0,
\end{equation}
where kernels $W^n$ were constructed from the adjacency matrices $A^n$ (cf.~\eqref{def-Wn}).

Convergence in the $\infty\to 1$ norm has strong implications for kernel operators
$$
\bW^n[f](x) \doteq \int_0^1 W^n(x,y) f(y) dy,\; n\in\N;\quad \bW [f](x) \doteq \int_0^1 W(x,y) f(y) dy.
$$
Operators $\bW^n$ and $\bW$ are symmetric Hilbert-Schmidt operators \cite{Young-book}.
Consequently, the spectrum
of each of them consists of a countable sequence of eigenvalues with a single accumulation point at
$0$:
\begin{align}\label{eig-Wn}
  \bW^n: \quad& \lambda^n_{-1}\le \lambda_{-2}^n\le \dots\le \lambda^n_0=0\le \dots\le \lambda_2^n\le \lambda^n_1,\\
  \label{eig-W}
 \bW: \quad& \lambda_{-1}\le \lambda_{-2}\le \dots\le \lambda_0=0\le \dots\le \lambda_2\le \lambda_1.
\end{align}
\begin{remark}\label{rem.eigenvalues}
  Equations \eqref{eig-Wn} and \eqref{eig-W} cover the general case when the spectra of $\bW^n$
and $\bW$ contain both positive and negative eigenvalues. If, for instance,
the spectrum of $\bW^n$ contains only nonnegative eigenvalues, all nonpositive eigenvalues
in \eqref{eig-Wn} are set to zero.
\end{remark}

Convergence in $\infty\to 1$ implies convergence of the corresponding eigenvalues of $\bW^n$
and $\bW$:
$$
\lim_{n\to\infty} \lambda^n_k=\lambda^k,\quad k\in\Z.
$$
Moreover, the corresponding eigenspaces also converge (see \cite[Lemma~3.3]{GhaMed2025}
for a precise statement).

We conclude this section with a brief discussion of the implications of the spectral convergence of
$\bW^n$ for signal processing. Recall that $W^n$ represents $A^n$, the adjacency matrix of
$\Gamma^n$ (cf.~\eqref{def-Wn}). The eigenspaces of the kernel operators $\bW^n$ are
directly related to those of $A^n$. Consequently, for sufficiently large $n$, the eigenfunctions
of the limiting operator $\bW$ can be used to approximate the Fourier coefficients in
\eqref{F-expansion}.

This approach offers multiple benefits. For random graph models, it eliminates the need to
compute eigenvectors of the adjacency matrix for every realization of the graph. Moreover,
as we will see, the eigenvectors obtained from the limiting graphon are smoother than
those from random adjacency matrices. These properties are central to GSP, which
uses graph limits for signal processing \cite{Ruiz2021}.
In the next section, we apply this technique to the stochastic block model for neuronal
connectivity.

\section{Stochastic block network} \label{sec.block}
\setcounter{equation}{0}

In this section, we formulate and analyze stochastic block network, which will be used in the
computational model in the next section.

Let
$$
B^n=C\otimes K^n,
$$
where 
$$
C=\begin{pmatrix} 1-2\alpha & \alpha & \alpha & 0\\
  \alpha & 1-2\alpha & 0 & \alpha \\
  \alpha & 0 & 1-2\alpha & \alpha\\
  0 & \alpha & \alpha & 1-2\alpha
  \end{pmatrix},\qquad 0<\alpha<1/2,
$$
and $K^n$ denotes an $n\times n$ adjacency matrix of a complete graph
$$
K^n=\begin{pmatrix} 0 & 1 & 1& \dots & 1& 1\\
  1 & 0 & 1 & \dots & 1& 1\\
  1 & 1& 1& \dots & 1& 0
\end{pmatrix}.
$$

\begin{figure}[h]
%\nopagenumber
%\renewcommand{\baselinestretch}{1.0}
\hfill
\begin{center}
\includegraphics[width=0.9\textwidth]{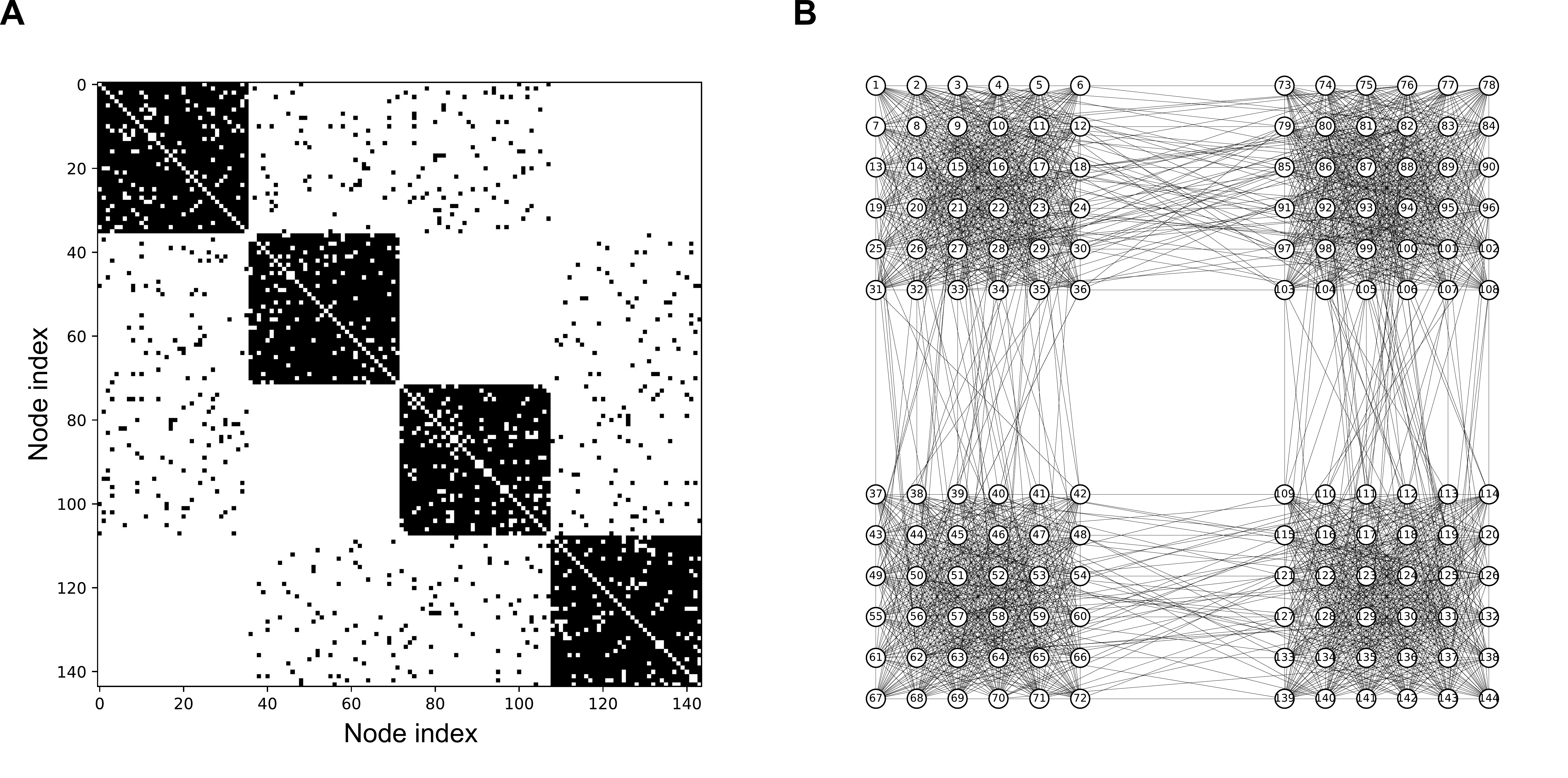}
\end{center}
\caption{Stochastic block network ($\alpha=0.05$). \textbf{(A)}\; A pixel picture of the adjacency matrix for a realization
  of the stochastic block model \eqref{Bernoulli}. \textbf{(B)}\;
A schematic representation of the stochastic block architecture.}
\label{f.block}
\end{figure}

Next, we define a $4n\times 4n$ random symmetric matrix $A^n$, whose entries are distributed
as Bernoulli random variables
\begin{equation}\label{Bernoulli}
A^n_{ij}=A^n_{ij} \sim \operatorname{Bernoulli}~(B^n_{ij}).
\end{equation}
$A^n$ is the adjacency matrix of a stochastic block network shown in Fig.~\ref{f.block}.
It has $4$ blocks with the density of connections inside blocks equal to $1-2\alpha$. In addition,
blocks $1$ and $4$ are connected with each of the blocks $2$ and $3$ with density $\alpha$.
The block architecture is chosen to match the biological network engineered in \cite{Sumi2023}.
Random
connectivity is used because the exact connectivity in the experimental network is not known.
Unless otherwise stated, we use $\alpha = 0.05$ throughout this paper; if a different value is used,
it will be explicitly specified. 
This network architecture will be
implemented in the computational model in the next section.

Following the scheme explained in the previous section, we represent the adjacency matrices
of $\Gamma^n$ by graphons. To this end, we define
\begin{equation}\label{graphons}
W^n(x,y)=\sum_{i,j=1}^{4n} A_{ij}^n\1_{I^n_i}(x) 1_{I^n_j}(y), 
\end{equation}
where $I_{n,i}, \; i\in [4n],$ are defined in \eqref{def-I} with $a_n:=4n$.
Next, we identify the graph limit
\begin{equation}\label{g-limit}
 %  W(x,y)=\sum_{i,j=1}^{4} \sum_{k,l=1}^n C_{ij}\1_{\left[\frac{(i-1)n+k-1}{4n}, \frac{(i-1)n+k}{4n}\right)  } (x)
 %  1_{\left[\frac{(j-1)n+l-1}{4n}, \frac{(j-1)n+l}{n}\right)  } (y)
W(x,y)=\sum_{i,j=1}^{4} C_{ij}\1_{I_i}(x)\1_j(y),  
\end{equation}
where $ I_i$ stands for $\left[\frac{i-1}{4}, \frac{i}{4}\right),\; i\in [4]$.
The corresponding kernel operator is given by
\begin{equation}\label{kernel}
  \bW[f](x)=\int_0^1W(x,y) f(y)dy.  
\end{equation}

Our next goal is to compute the eigenvalues and the corresponding eigenfunctions
of $\bW$. To this end, we first find the eigenvalues of $C$. In the table below,
for each eigenvalue we provide a basis of the corresponding eigenspace.

\begin{center}
    \begin{tabular}{ll} 
      \hline
       & \\
      $\lambda_1=1$ \qquad\qquad&\qquad\qquad
        $v_1=\begin{pmatrix}  \frac{1}{2}\\ \frac{1}{2}\\ \frac{1}{2}\\ \frac{1}{2} \end{pmatrix}$\\
      $\lambda_2=\lambda_3=1-2\alpha$ \qquad\qquad&\qquad\qquad
      $v_2=\begin{pmatrix}  0\\ \frac{1}{\sqrt{2}}\\ \frac{-1}{\sqrt{2}}\\ 0 \end{pmatrix}$,
      $v_3=\begin{pmatrix}  \frac{1}{\sqrt{2}}\\ 0\\ 0\\ \frac{-1}{\sqrt{2}} \end{pmatrix}$,\\
       $\lambda_4=1-4\alpha$ \qquad\qquad &\qquad\qquad
                                            $v_4=\begin{pmatrix}  \frac{1}{2}\\ \frac{-1}{2}\\ \frac{-1}{2}\\ \frac{1}{2} \end{pmatrix}$.\\
      & \\
        \hline
    \end{tabular} 
  \end{center}
  The eigenvectors $v_1, v_2, v_3,$ and $v_4$ form an orthonormal basis in $\R^4$.

\begin{figure}[h]
%\nopagenumber
%\renewcommand{\baselinestretch}{1.0}
\hfill
\begin{center}
\includegraphics[width=0.9\textwidth]{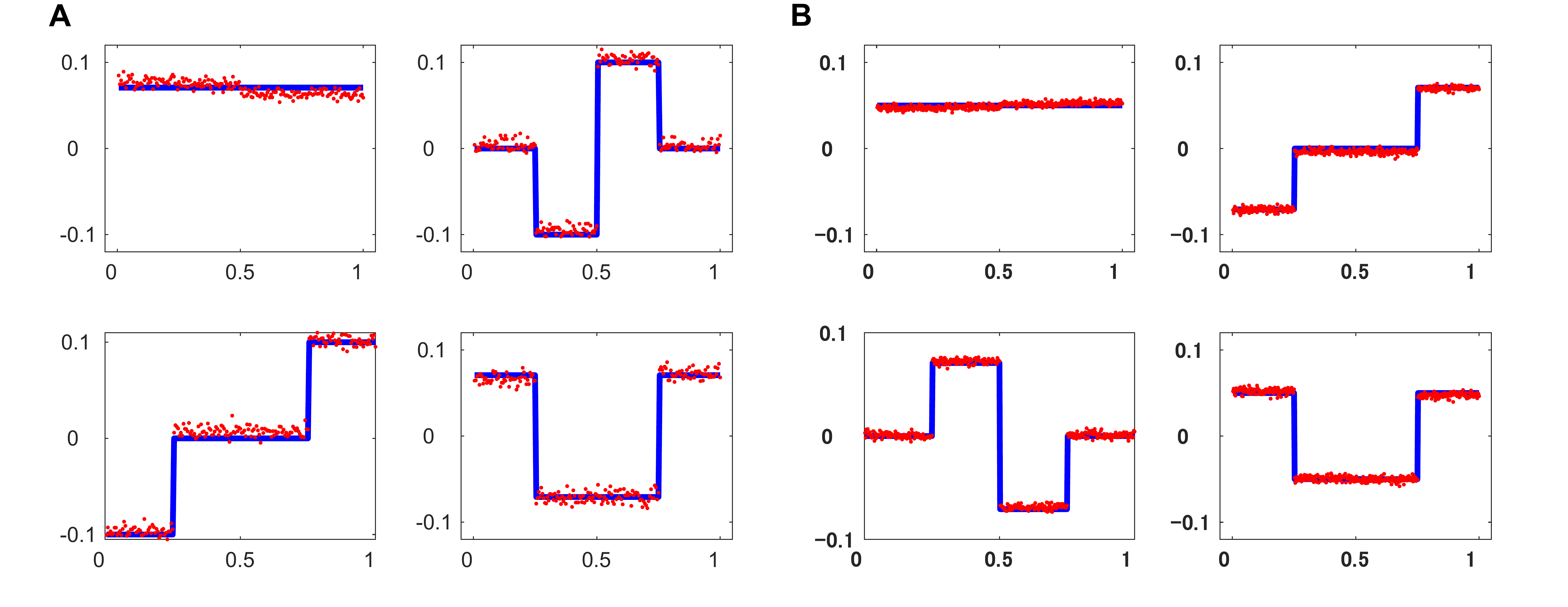}
\end{center}
\caption{The eigenbasis of the subspace corresponding to the nonzero eigenvalues of $\bW$
  (plotted in solid blue line)
  and the corresponding eigenfunctions of  random realization of $\bW^n$ (red dots). Displays
  \textbf{(A)} and \textbf{(B)}, corresponding to $n=50$ and $n=100$ respectively, illustrate
  convergence of the eigenspaces of $\bW^n$ to the corresponding eigenspaces of $\bW$.
}
\label{f.gphon-eigf}
\end{figure}

With the eigenvalues and the eigenvectors of $C$ in hand, it easy to deduce the eigenvalues
of $\bW$. Below, we list nonzero eigenvalues followed by the corresponding eigenfunctions
on $[0,1]$.

  \begin{center}
    \begin{tabular}{ll} 
      \hline
       & \\
      $\lambda_1=1/4$ \qquad\qquad&\qquad\qquad
        $\phi_1(x)=1 $\\
      $\lambda_2=\lambda_3=(1-2\alpha)/4$ \qquad\qquad&\qquad\qquad
      $\phi_2= \sum_{j=1}^4 v_{2j} \1_{I_j}(x)$, \; $\phi_3= \sum_{j=1}^4 v_{3j} \1_{I_j}(x)$
                \\
       $\lambda_4=(1-4\alpha)/4$ \qquad\qquad &\qquad\qquad
      $\phi_4= \sum_{j=1}^4 v_{4j} \1_{I_j}(x)$.\\                                        
      & \\
        \hline
    \end{tabular} 
  \end{center}

  In addition, there is a zero eigenvalue $\lambda_0=0$. The corresponding eigenspace is given by
$$
V_0=\left\{ f\in L^2([0,1]):\; \int_{I_i} f(x)dx=0, \; i=1,2,3,4\right\}.
$$

We are now ready to demonstrate the advantages of using graphons for analyzing signals
on $\Gamma^n$. Figure~\ref{f.gphon-eigf} compares the eigenfunctions of the kernel operator
$\bW$ with those of the stochastic block network. Panels A and B show that,
as $n$ increases, the eigenfunctions of the stochastic network converge to those of $\bW$,
so the latter provide accurate approximations.  

A key benefit is that the eigenfunctions of $\bW$ are piecewise constant and (piecewise) smooth, whereas
those of $\bW^n$ are highly irregular due to randomness in each realization of the graph.
Thus, for large $n$, one not only avoids loss of information by using the continuum limit,
but in fact gains by working with cleaner, noise-free eigenfunctions that capture the essential
block structure of the network.

We conclude this section with a discussion of the geometric meaning of the Fourier coefficients
with respect to the eigenbasis of $\bW$.  
Let $f$ be a nonnegative function on $[0,1]$ with $\|f\|_{L^1([0,1])}=1$. It can be decomposed
into a Fourier series with respect to the eigenfunctions of $\bW$:  
\begin{equation}\label{f-expand}
f=\sum_{i=1}^4 (f,\phi_i)\,\phi_i + P_{V_0}f,
\end{equation}
where $P_{V_0}$ denotes the orthogonal projection onto the infinite-dimensional eigensubspace
$V_0$ corresponding to the zero eigenvalue.

The first four modes in \eqref{f-expand} correspond to the nonzero eigenvalues of $\bW$, which
coincide with the nonzero eigenvalues of $C$. The eigenfunctions $\phi_i, \, i\in [4],$
inherit their structure from the eigenvectors $v_i, \, i\in [4],$ of $C$, thus, encoding the
block architecture of the network. This relation gives a clear geometric interpretation of the
leading Fourier coefficients in \eqref{f-expand}.

Specifically,
$$
(f,\phi_1)=\int_0^1 f(x)\,dx = 1
$$
measures the total “mass” of the signal. Next,
$$
(f,\phi_2)=\int_{I_2} f(x)\,dx - \int_{I_3} f(x)\,dx
$$
quantifies the difference between the projections of the signal on the second and third clusters.
Similarly, the third and fourth coefficients,
$$
(f,\phi_3)=\int_{I_1} f(x)\,dx - \int_{I_4} f(x)\,dx,
\quad
(f,\phi_4)=\int_{I_1\cup I_4} f(x)\,dx - \int_{I_2\cup I_3} f(x)\,dx,
$$
encode differences across other cluster partitions.

Thus, although the Fourier expansion of $f$ involves infinitely many terms, the coefficients
$(f,\phi_i), \; i\in\{2,3,4\},$ play a central role for signal processing: they capture how the
signal aligns with the modular architecture of the network. This property will be central to our
analysis of the SIP in the next section.

\section{The integrate-and-fire network model}
\label{sec.spiking}
\setcounter{equation}{0}

In this section, we formulate the integrate-and-fire neuron model, which will be used
as a testbed for our graphon-based approach to the SIP addressed in the next section.

Consistent with the network architecture explained in the previous section, we consider
a network of $4n$ neurons organized into distinct blocks and interacting via excitatory
synapses.
The membrane potential on neuron \(V_i(t)\) of neuron $i\in [4n]$
evolves according a leaky integrate-and-fire neuron model:
\begin{equation}\label{iaf}
\tau_{\rm mem}\,\frac{dV_i}{dt}
= E_L - V_i(t) + R_{\rm in}\,I_{\rm tot}(t),
\quad
\tau_{\rm mem}=20\ \mathrm{ms},\;
E_L=-74\ \mathrm{mV},\;
R_{\rm in}=4.0\times10^4\ \mathrm{k\Omega}.
\end{equation}
When \(V_i\) reaches the threshold \(V_{\rm th}=-54\ \mathrm{mV}\), a spike is emitted,
\(V_i\) is reset to \(V_{\rm reset}=-60\ \mathrm{mV}\).

The total input current consists of synaptic, calcium-dependent K\(^+\),
refractory, and input terms \cite{French2006, Yamamoto2016, Ishikawa2024, Sato2024}:
\begin{equation*}
I_{\rm tot}(t)
= \sum_{j}I_j(t)
\;+\;I_{K({\rm Ca})}(t)
\;+\;I_{\rm ref}(t)
\;+\;I_{\rm in}(t),
\end{equation*}
where \(I_{{\rm in}}\) is input current, and
the details of \(I_{K({\rm Ca})}\) and \(I_{\rm ref}\) are given in \cite{Yamamoto2016}.

The synaptic currents \(I_j(t)\) are modeled in terms of conductance \(g_j(t)\):
\begin{align*}
I_j(t) &= g_j(t)\,\bigl(E_{\rm syn} - V_i(t)\bigr), \\[6pt]
\frac{d\,g_j(t)}{dt}
       &= -\frac{g_j(t)}{\tau}
          + \frac{g_{\rm max}}{S_j}\sum_{i} A^n_{ji}R_i\bigl(\sum_{k}\delta\bigl(t - (t_i^k + d_{ij})\bigr)
          + \sum_{p}\delta\bigl(t - t_i^p\bigr)\bigr),
\end{align*}
with
\(E_{\rm syn}=0\ \mathrm{mV}\), \(g_{\rm max}=3.5 \times 10^{-4}\
\mathrm{mS}\), \(\tau=5.0\ \mathrm{ms}\). \(A^n_{ji}\) is
defined as in Section~\ref{sec.block}, \(S_{j}=\Sigma_iA^n_{ji}\) is a scaling factor. The presynaptic vesicle resource \(R_i\) is depressed at spikes (\(R_i \leftarrow \beta R_i\)) and otherwise relaxes towards unity with time constant \(\tau_R\) via \(\tau_R\,\dot{R}=1-R\), with \(\beta=0.8\) and \(\tau_R=2.0\times10^4\ \mathrm{ms}\).
The conduction delay \(d_{ij}\) is fixed at 2.8 ms,
\(t_i^k\) denotes the time of the \(k\) th spike of the neuron \(i\), and \(t_i^p\) denotes
the time of 1 Hz Poisson spikes used to model spontaneous synaptic noise.

\begin{figure}[h]
%\nopagenumber
%\renewcommand{\baselinestretch}{1.0}
\hfill
\begin{center}
\includegraphics[width=0.7\textwidth]{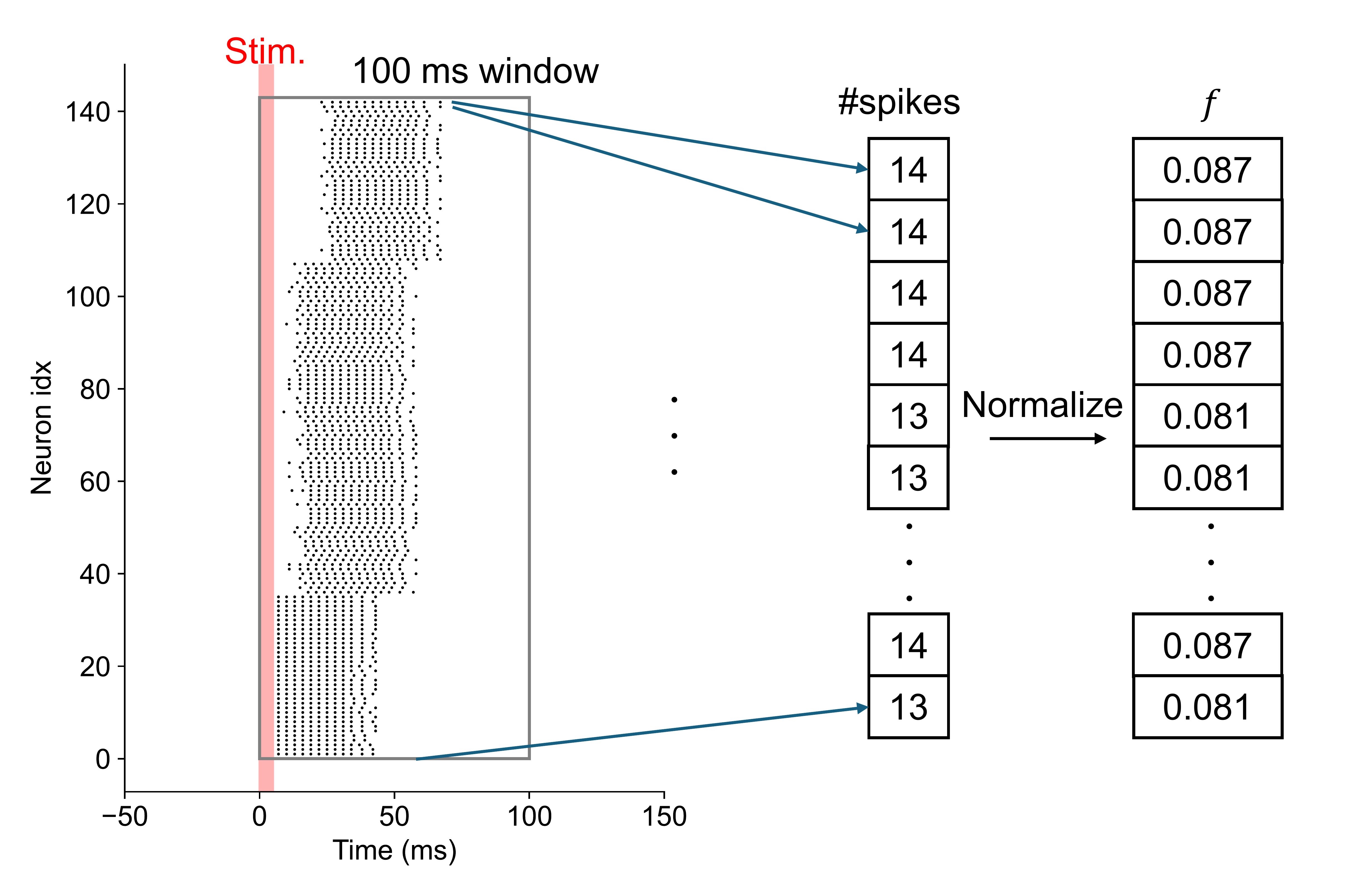}
\end{center}
\caption{Extraction of the signal from neural responses. Neural activity was represented as binary spike trains, taking a value of 1 at spike times (shown as black dots in the plot) and 0 otherwise. Responses to a 5-ms stimulation pulse were then integrated within a 100-ms time window and subsequently normalized such that their L2 norm equaled 1.}
\label{Fig:figure3}
\end{figure}

We conclude this section by illustrating the modeling equations with a set of representative
network simulations in response to a stimulus \(s\). At this stage, the exact form of the
stimulus is not important; the precise stimulation protocols used in the SIP will be introduced
in the next section. Our immediate goal is to clarify how the signal
\(f \in L(\R^{4n}, \R^+)\) is constructed from the neural dynamics generated by
\eqref{iaf}.  

Here, \(f\) represents the spiking network response, quantified as the spike count within
$0–100$ ms after stimulus onset. In these examples, \(s\) is delivered as a $5$-ms pulse to
\(I_{\text{in}}\). The mapping from neural activity to the signal \(f\) is illustrated in
Fig.~\ref{Fig:figure3}.  

This construction provides the bridge from microscopic neural dynamics to the macroscopic
signals that form the input to GnSP. In the SIP, the task is to infer
the identity of \(s\) based solely on the observed \(f\). The GnSP framework will operate
on these representations, using spectral projections to extract low-dimensional, robust
features that reflect the underlying network architecture. 

\section{The signal identification problem}
\label{sec.sip}
\setcounter{equation}{0}

In this section, we address the SIP for the integrate-and-fire network \eqref{iaf}.  
Roughly speaking, the goal is to identify which stimulus from the set
$s \in \{s_1, s_2, s_3\}$ was applied to the network based on the observed output $f$.  

We begin with a simpler variant in which the stimulus is localized to a single cluster.  
This setting allows us to clearly demonstrate the mechanism of GSP in a straightforward and
intuitive way. Next, we consider the more challenging case where the stimulus can be
distributed across multiple clusters. We show that our method remains effective even in
this more complex scenario.  

These computational experiments provide insight into the workings of GnSP. Unlike GSP on a
finite graph, which is sensitive to stochastic variations across network realizations, the
graphon formulation provides robustness to the “noise” introduced by individual random
graph realizations and finite-size effects.

\subsection{Two-cluster stimulation}

\begin{figure}[h]
%\nopagenumber
%\renewcommand{\baselinestretch}{1.0}
\hfill
\begin{center}
\includegraphics[width=0.9\textwidth]{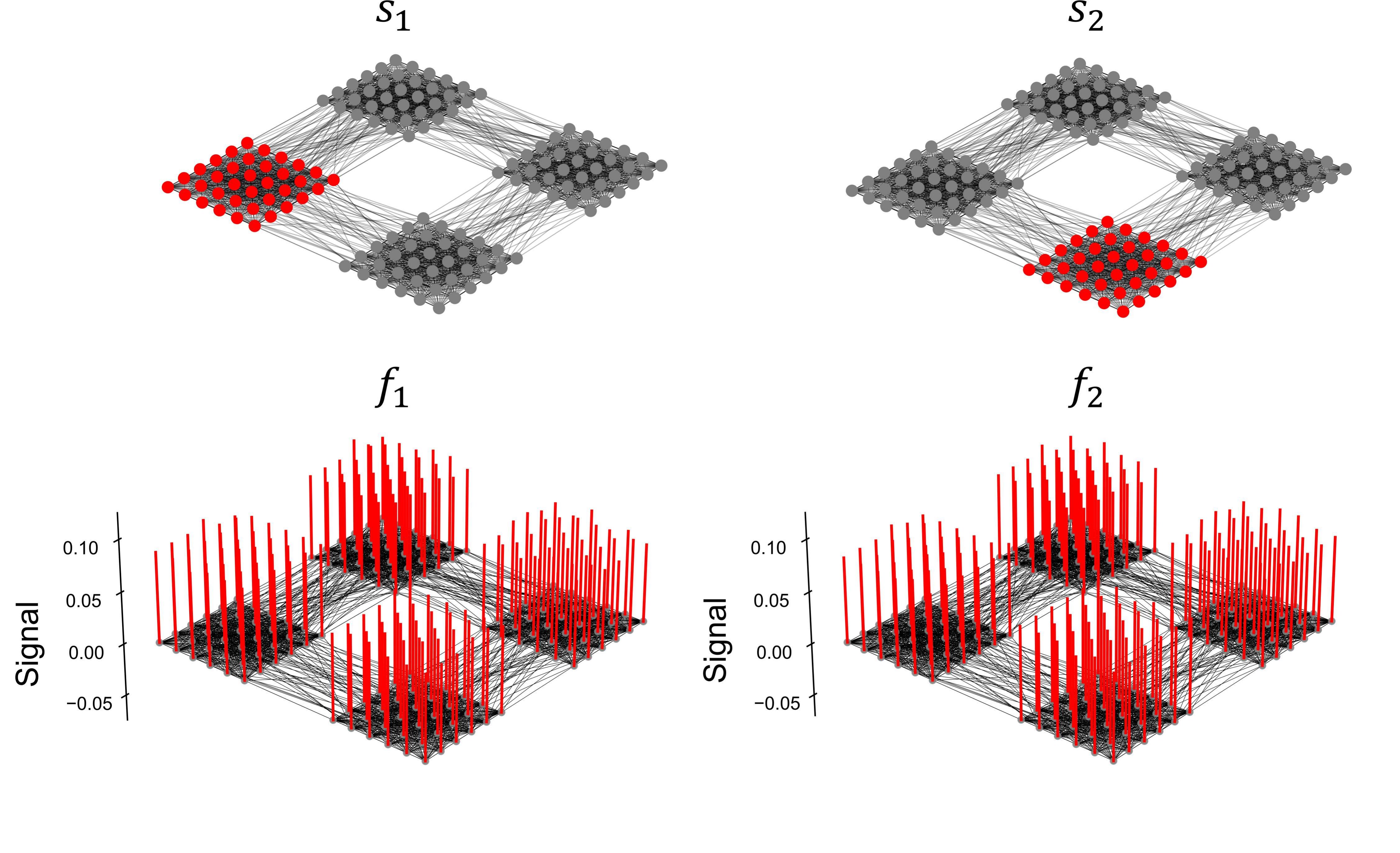}
\end{center}
\caption{Two-cluster stimuli (top) and their representative responses (bottom).}
\label{Fig:figure4}
\end{figure}

\begin{figure}[h]
%\nopagenumber
%\renewcommand{\baselinestretch}{1.0}
\hfill
\begin{center}
\includegraphics[width=0.9\textwidth]{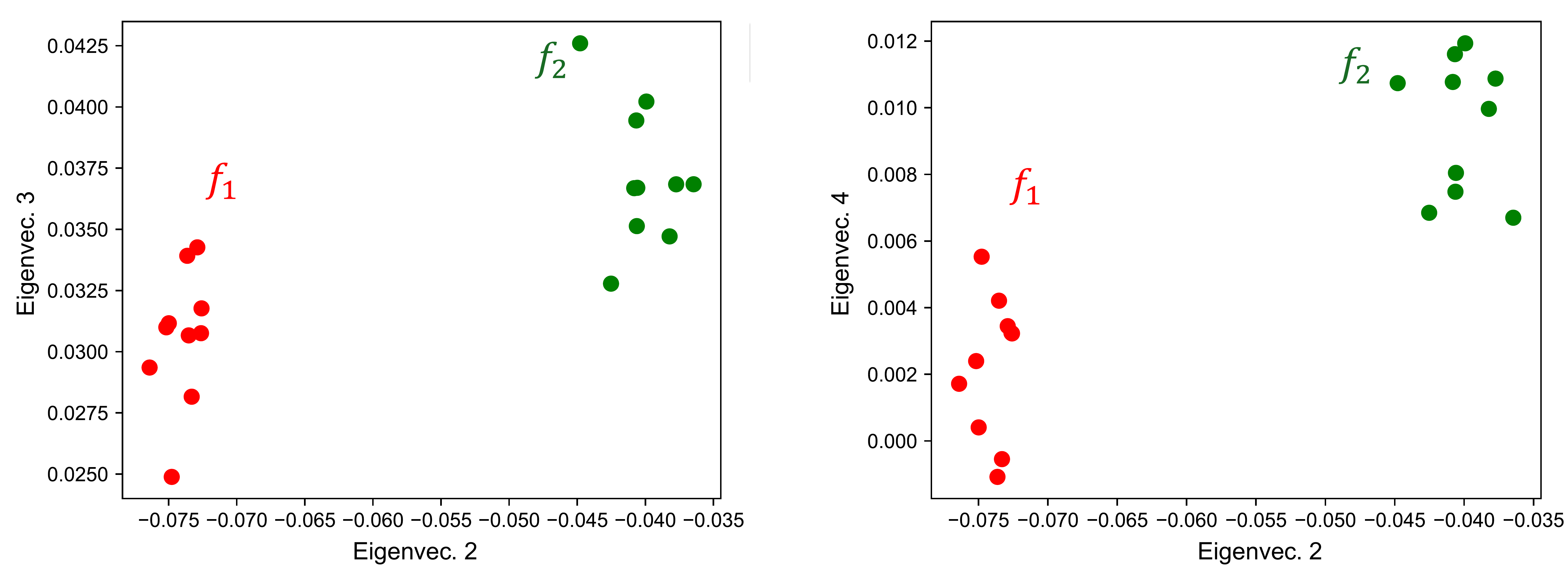}
\end{center}
\caption{Projection map onto eigenvector 2 and 3 (left) and eigenvector 2 and 4 (right). Red denotes $f_1$ and green $f_2$.}
\label{Fig:figure5}
\end{figure}

In this subsection, we consider the case when stimuli of two types are used:
Stimuli \(s_1\) and \(s_2\) are applied  to clusters $1$ and $2$, respectively
(see Fig.~\ref{Fig:figure4}).
Although
the stimuli are cluster-specific, the corresponding neural responses \(f_1\) and \(f_2\) propagate
throughout
the network, making them difficult to distinguish empirically.
Using GSP, we projected each
response \(f\) onto the eigenvectors (graph Fourier transform), yielding the maps in
Fig.~\ref{Fig:figure5}. In this spectral domain, \(f_1\) and \(f_2\) formed clearly separated clusters,
indicating that the projection improved discriminability.

\begin{figure}[h]
%\nopagenumber
%\renewcommand{\baselinestretch}{1.0}
\hfill
\begin{center}
\includegraphics[width=0.9\textwidth]{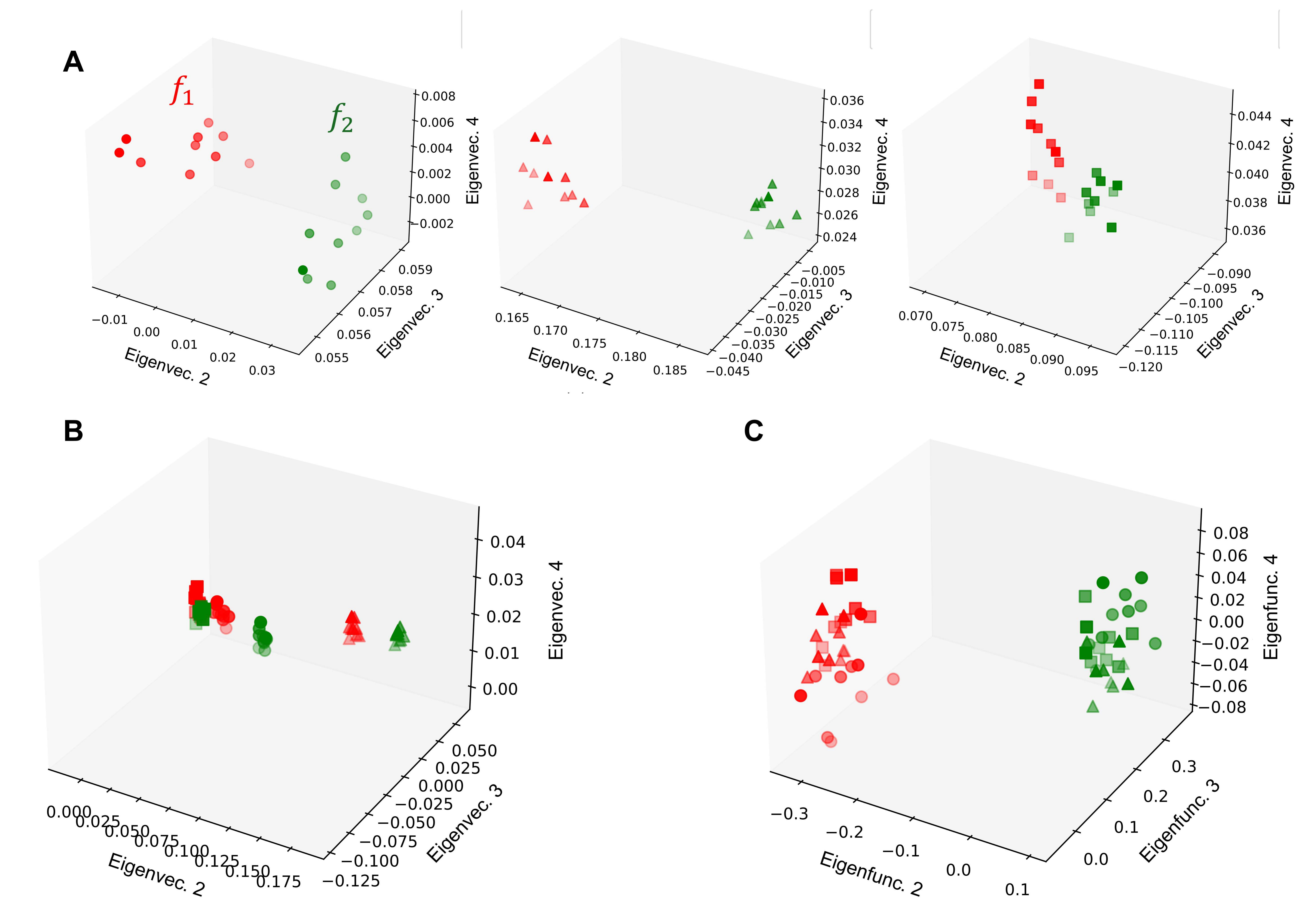}
\end{center}
\caption{Projection onto graph eigenvectors vs.\ graphon eigenfunctions.
\textbf{(A)} Three independent trials projected onto the graph eigenvectors; coordinates vary across trials due to stochasticity.
\textbf{(B)} The three trials overlaid, highlighting trial-to-trial scatter under the finite-graph basis.
\textbf{(C)} Projection onto the graphon eigenfunctions; coordinates are stable across trials and essentially insensitive to stochasticity.}
\label{Fig:figure6}
\end{figure}

However, this separation depends on the specific adjacency matrix \(A^n\) of each sampled graph.
Because
the eigenbasis varies with \(A^n\), finite-size fluctuations cause trial-to-trial variability in the
projected coordinates (Fig.~\ref{Fig:figure6}A), and overlaying trials produces misaligned maps
(Fig.~\ref{Fig:figure6}B). In contrast, projection onto graphon eigenfunctions produces coordinates
that are consistent across trials and robust to stochastic variability, while preserving the separability
between \(s_1\) and \(s_2\) (Fig.~\ref{Fig:figure6}C). This result demonstrates that the graphon-based
approach provides a stable and generalizable representation of network responses across different realizations.

\subsection{Mixed cluster stimulation}
\label{sec:mixed}
\begin{figure}[h]
%\nopagenumber
%\renewcommand{\baselinestretch}{1.0}
\hfill
\begin{center}
\includegraphics[width=0.9\textwidth]{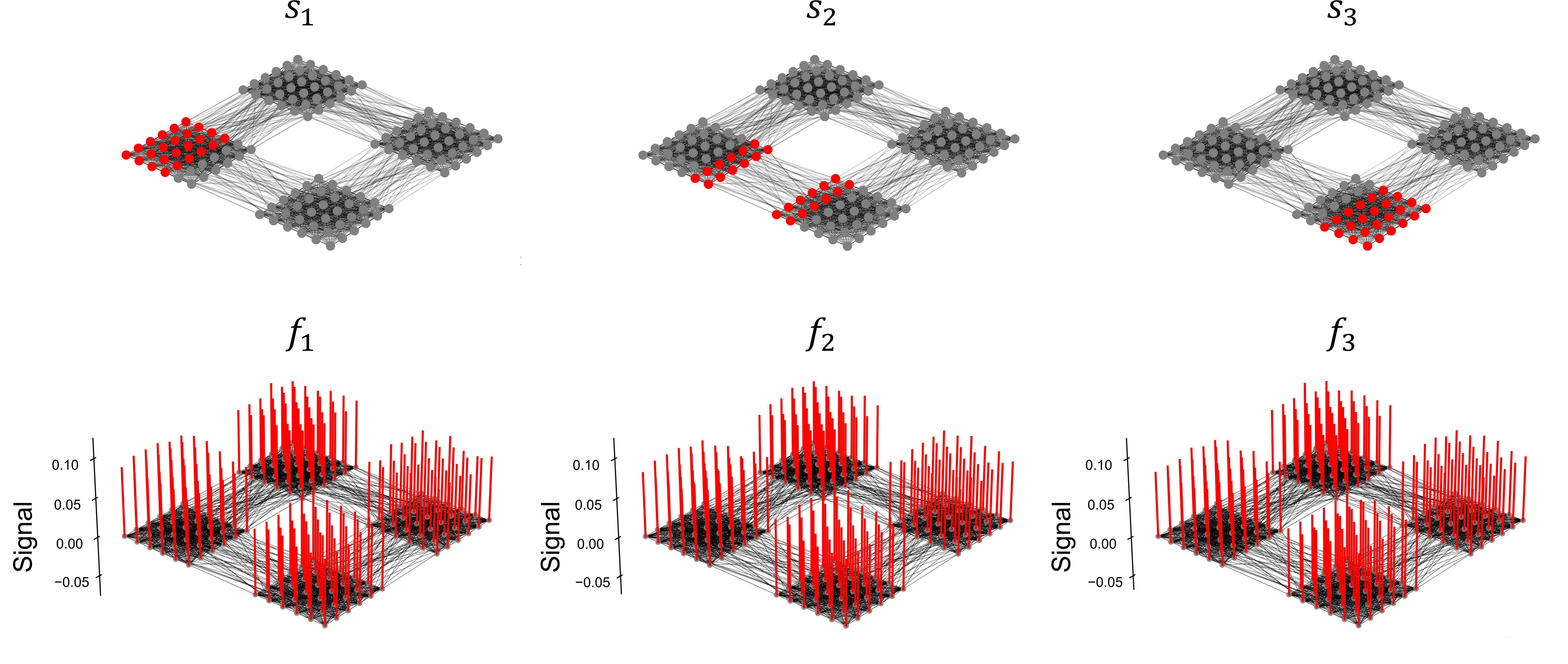}
\end{center}
\caption{Three stimuli (top) and their representative responses (bottom).}
\label{Fig:figure7}
\end{figure}

%Takuma will write
Next, we examine a scenario in which three distinct stimuli,
\(s_1\), \(s_2\), and \(s_3\), are applied in a mixed cluster manner
(Fig.~\ref{Fig:figure7}). Specifically, \(s_1\) targets two-thirds of the nodes
in cluster~1, \(s_2\) stimulates the remaining one-third of cluster~1 together
with one-third of cluster~2, and \(s_3\) targeted the remaining two-thirds of
cluster~2. This configuration produces a mixed cluster pattern for
\(s_2\), in which the stimulated nodes are drawn from both clusters.

\begin{figure}[h]
%\nopagenumber
%\renewcommand{\baselinestretch}{1.0}
\hfill
\begin{center}
\includegraphics[width=0.9\textwidth]{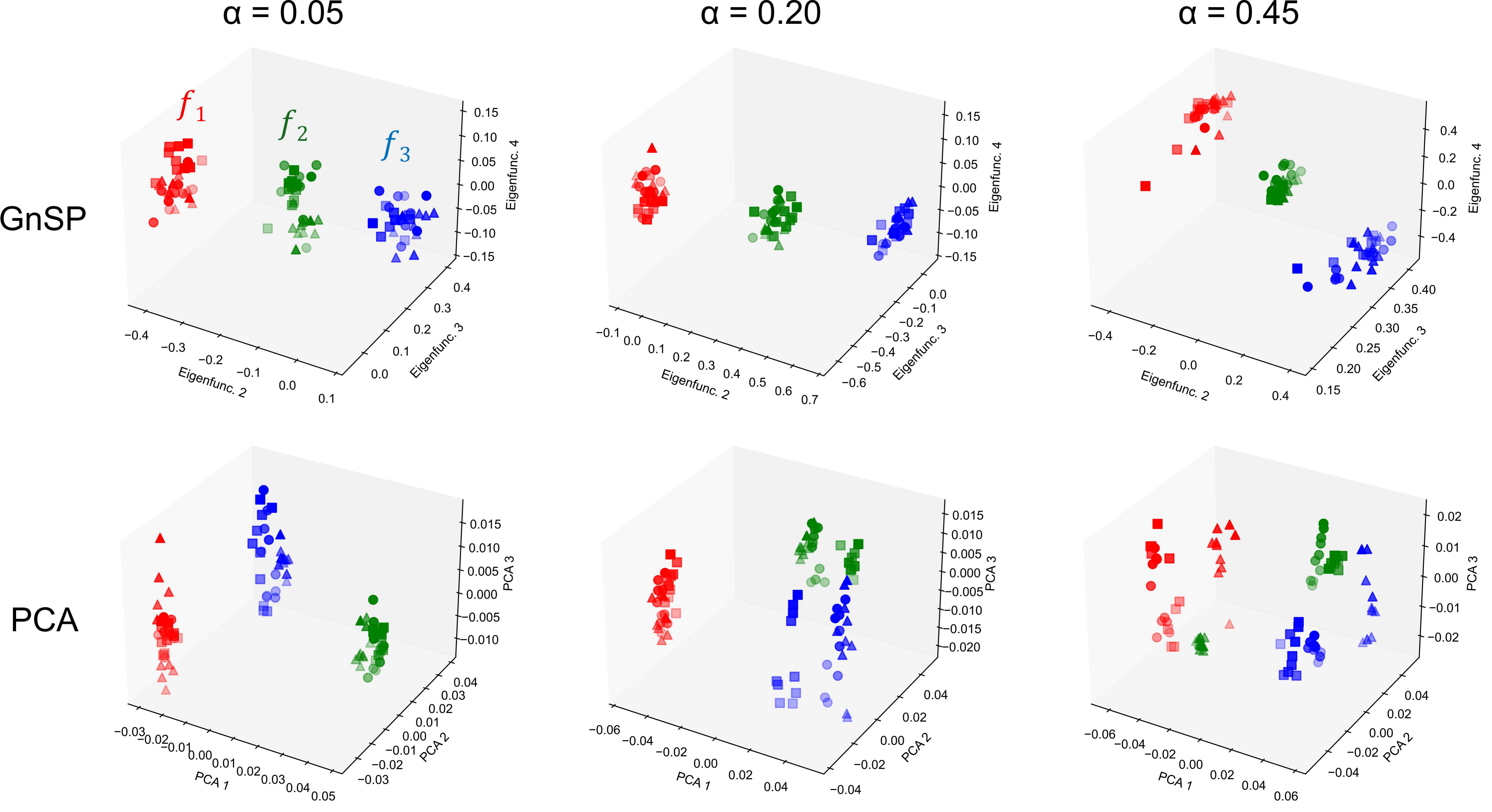}
\end{center}
\caption{Comparison of GnSP embeddings and PCA projections 
for neural response. 
Top row: GnSP mappings obtained for three values of the inter-block 
connectivity parameter $\alpha$ ($\alpha=0.05,\,0.20,\,0.45$). 
Bottom row: corresponding PCA projections for the same networks. Red denotes $f_1$, green $f_2$, and blue $f_3$. In the GnSP embeddings, A stimulus $s_2$ spanning two clusters evokes a response $f_2$ located between $f_1$ and $f_3$.}

\label{Fig:figure8}
\end{figure}

We projected the resulting responses \(f_1\), \(f_2\), and \(f_3\) onto the graphon
eigenfunctions (Fig.~\ref{Fig:figure8}). In this representation, each stimulus produced
a distinct and reproducible cluster in the low-dimensional map.
Interestingly, the response \(f_2\) to the mixed-cluster stimulus appeared in
an intermediate location between the \(f_1\) and \(f_3\) groups,
reflecting its mixed stimulation pattern. This spatial arrangement on the map illustrates
how graphon-based coordinates can capture graded relationships between
network responses arising from stimuli spanning both clusters.

We then examined the robustness of GnSP with respect to changes in the stochastic block
model parameter~$\alpha$, 
which controls the level of inter-block connectivity. 
The graphon-based embeddings remained stable across different values of~$\alpha$ 
($\alpha=0.05,\,0.20,\,0.45$), consistently producing three well-separated clusters
(Fig.~\ref{Fig:figure8} (Top)). 
For comparison, principal component analysis (PCA), a standard method for dimensionality
reduction, 
yielded representations in which the three clusters became progressively less distinct 
as~$\alpha$ increased, indicating reduced robustness to network stochasticity
(Fig.~\ref{Fig:figure8} (Bottom)). 
These results demonstrate that GnSP provides embeddings that are resilient 
to variability in the generative parameters of the network.

\section{Applying GnSP to experimental data}
\label{sec.experiment}
\begin{figure}[h]
%\nopagenumber
%\renewcommand{\baselinestretch}{1.0}
\hfill
\begin{center}
\includegraphics[width=0.9\textwidth]{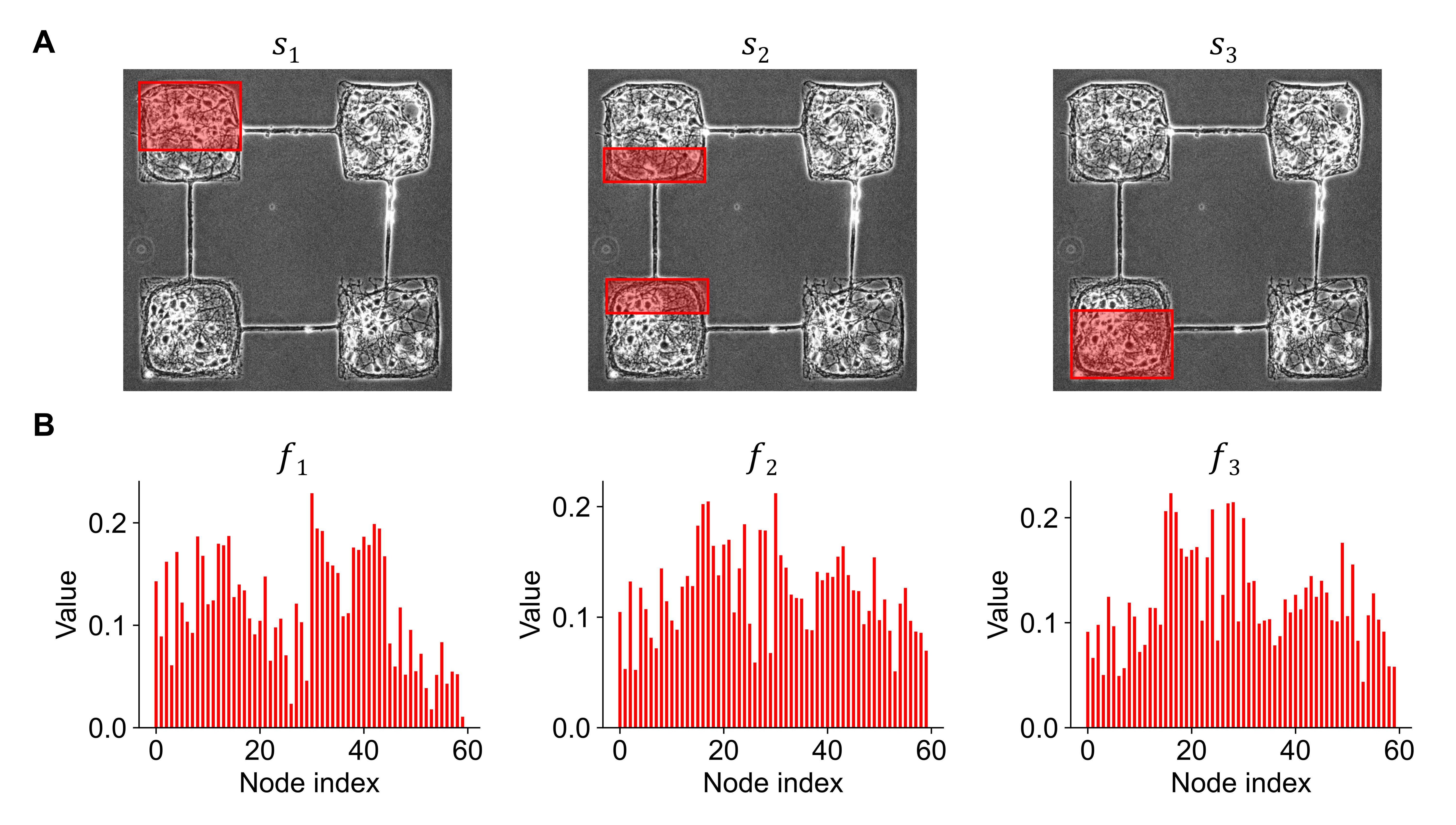}
\end{center}
\caption{\textbf{(A)} Three types of photostimulation patterns applied to the cultured neuronal network, and \textbf{(B)} the representative responses to each stimulus.}
\label{Fig:figure9}
\end{figure}

\setcounter{equation}{0}
To assess the applicability of the proposed graphon-based method to
real biological data, we analyzed the calcium-imaging dataset of
modular cultured neuronal networks available at
\href{https://zenodo.org/records/7792577}{zenodo.org/records/7792577}
\cite{Sumi2023Zenodo}. In that
study, networks were stimulated under three conditions (\(s_1\),
\(s_2\), \(s_3\))
comparable to those described in the previous sections (Fig.~\ref{Fig:figure9}).
Calcium fluorescence signals were interpreted as proportional to the
cumulative spike
count over a short time window, and each neural response \(f\) was
quantified as the fluorescence intensity measured 100~ms after
stimulus onset. Response vectors were normalized to have unit
L2 norm before further analysis.

\begin{figure}[h]
%\nopagenumber
%\renewcommand{\baselinestretch}{1.0}
\hfill
\begin{center}
\includegraphics[width=0.9\textwidth]{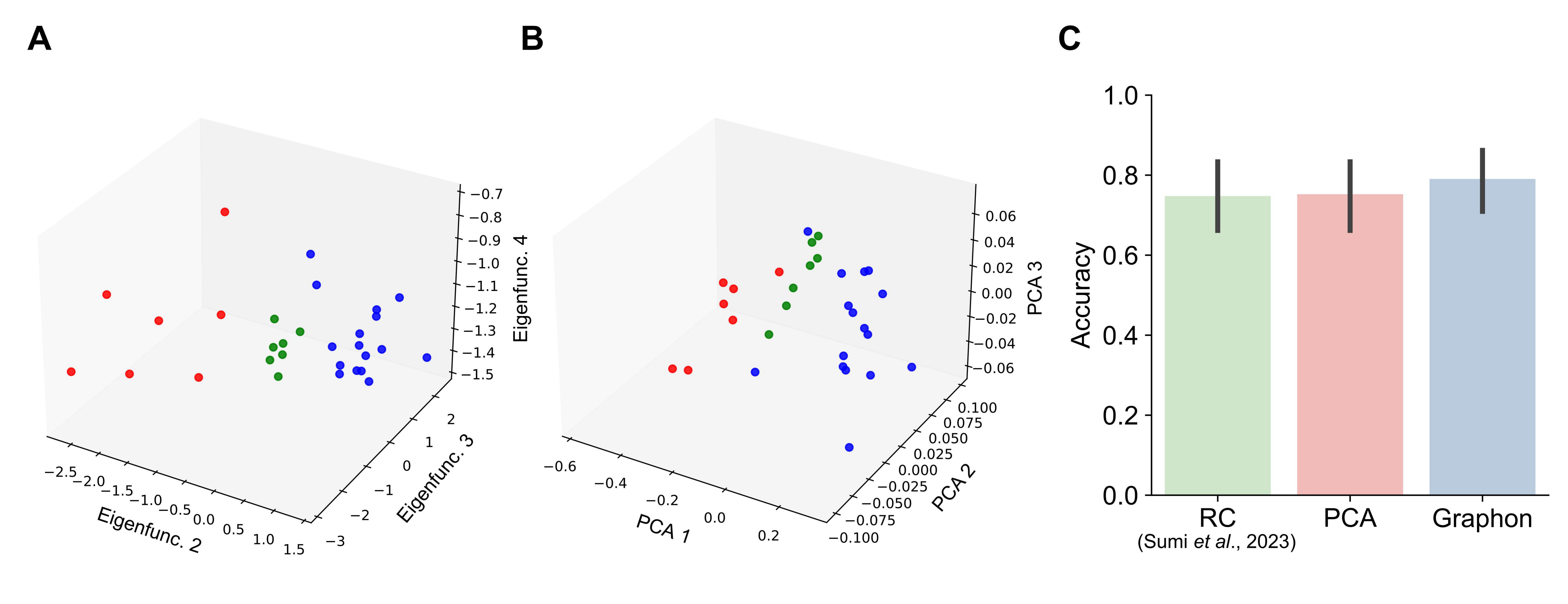}
\end{center}
\caption{\textbf{(A)} Three-dimensional mapping of responses from the biological neural network using Graphon eigenfunctions. \textbf{(B)} The same mapping obtained using PCA. \textbf{(C)} Bar plot comparing the accuracy of the stimulus regression task across RC (as used in previous work \cite{Sumi2023}), PCA, and Graphon. Error bars represent 95\% confidence intervals, with $n=21$.}
\label{Fig:figure10}
\end{figure}

We projected these responses into a low-dimensional space using GnSP (Fig.~\ref{Fig:figure10}A) and PCA (Fig.~\ref{Fig:figure10}B). Both methods produced separable and reproducible clusters corresponding to the three stimuli. Following the classification protocol of \cite{Sumi2023}, we applied ridge regression to four-dimensional projections from each method. The resulting classification accuracies were \(0.752\) for PCA and \(0.790\) for the graphon-based method, the latter exceeding the reservoir computing (RC) result reported in \cite{Sumi2023} (accuracy \(0.748\)) (Fig.~\ref{Fig:figure10}C).

Despite this numerical improvement, the 95\% confidence interval for the accuracy difference between GnSP and RC data (\([-0.0381, 0.129]\)) included zero, indicating that the gain was not statistically significant with the available \(n=21\) samples. The estimated effect size (\(0.213\)) suggests that detecting such a difference with 80\% power at a significance level of 0.05 would require approximately \(n \approx 173\) samples.

These results suggest that, although confirming them statistically will require substantially larger datasets, the graphon-based method may offer improved discriminability. We anticipate that applying this approach to a wider range of experimental systems, combined with larger datasets, will allow its advantages to be demonstrated conclusively and further advance the study of complex neural dynamics.

\section{Beyond stochastic block model}
\label{sec.generalization}

The stochastic block architecture adopted for numerical experiments in Section~\ref{sec.sip} was inspired by the
experimental setting in \cite{Sumi2023}. The analysis of the computational and experimental data clearly demonstrates the benefits of the GnSP for the signal identification problem. It is natural to ask, however, how well this approach extends to more general network architectures. While the stochastic block model may appear as a very special type of network organization, any dense network can be approximated by a stochastic block one, thanks to the Weak Regularity Lemma. Specifically, any graphon can be approximated by a step graphon in the cut norm, with the number of steps depending only on the desired accuracy (see Lemma~9.9 in \cite{Lovasz-book}).

In this section, we show that our approach extends naturally to small-world networks, another common motif in complex networks \cite{Bassett&Bullmore2006, Sporns2004}.
We apply the GnSP framework developed in the previous sections to small-world graphons \cite{Medvedev2014}
and then extend the analysis to networks with a larger number of blocks to evaluate its robustness and scalability across complex networks.

\subsection{Small-world graphon}
Small-world networks provide a canonical example of continuous and distance-dependent connectivity that extends beyond the discrete community structure assumed in the four-block stochastic model.
To examine the generality of the GnSP framework, we first considered a small-world graphon defined as
\begin{equation}
W(x,y) =
\begin{cases}
1 - p, & \min(|x - y|, 1 - |x - y|) \leq r,\\
p, & \text{otherwise}.
\end{cases}
\label{eq.swgraphon}
\end{equation}
The parameter $r$ determines the range of local connections, and $p=0.05$ controls the probability of random long-range links, generating small-world connectivity.
Applying the GnSP framework to this kernel allows us to examine whether the proposed graphon-based embedding remains effective across different network structures.

\begin{figure}[h]
%\nopagenumber
%\renewcommand{\baselinestretch}{1.0}
\hfill
\begin{center}
\includegraphics[width=0.9\textwidth]{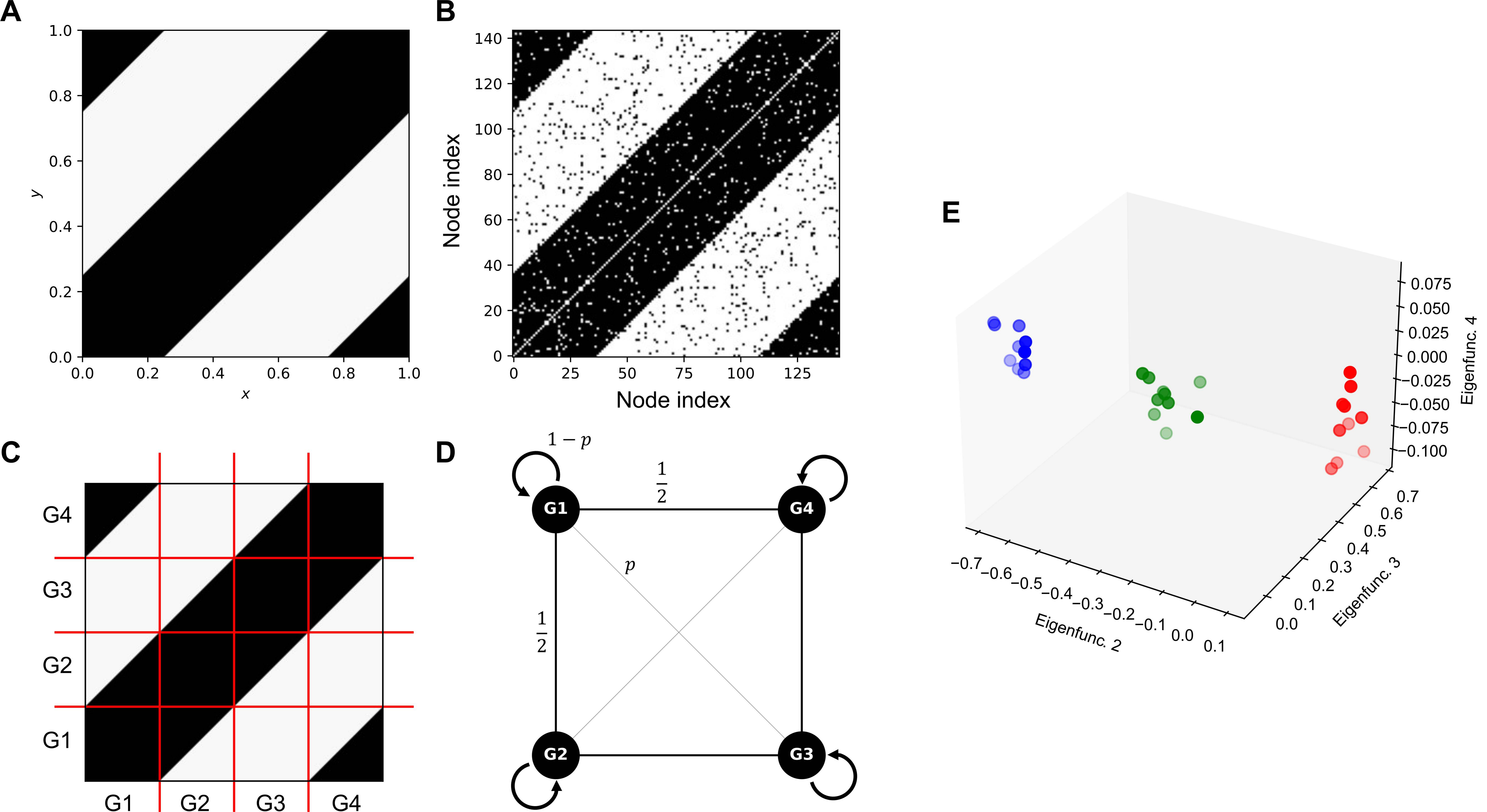}
\end{center}
\caption{\textbf{(A)} Small-world graphon with parameters \(r = 1/4\) and \(p = 0.05\). \textbf{(B)} Example of an adjacency matrix sampled from the small-world graphon. \textbf{(C)} Partition of the graphon domain into four groups. \textbf{(D)} The resulting four-block structure exhibits the same modular organization as the stochastic block model (SBM) graphon used in previous sections. \textbf{(E)} Signal identification problem (SIP) for network activity generated by 
integrate-and-fire simulations on small-world graphs.
Signals were projected onto the eigenfunctions of the four-block approximation of the small-world graphon.}
\label{Fig:figure11}
\end{figure}

We first fixed the local connection range at $r = 1/4$. 
The corresponding small-world graphon is shown in Fig.~\ref{Fig:figure11}A, and an example of a sampled adjacency matrix is presented in Fig.~\ref{Fig:figure11}B. As illustrated in Fig.~\ref{Fig:figure11}C, this small-world graphon can be partitioned into four groups. Interestingly, this partition yields a four-block structure whose network representation (Fig.~\ref{Fig:figure11}D) exhibits the same modular organization as the stochastic block model (SBM) graphon used earlier.
In the following analysis, we show that this correspondence extends to the spectral domain, where the eigenfunctions of the four-block approximation of the small-world graphon align with those of the stochastic block model.

To demonstrate this spectral correspondence, we derive the eigenstructure of the four-block approximation obtained by partitioning the small-world graphon.
The resulting block-averaged connectivity matrix is
$$
B^n=C\otimes K^n,
$$
where 
$$
C=\begin{pmatrix} 1-p & \frac{1}{2} & p & \frac{1}{2}\\
  \frac{1}{2} & 1-p & \frac{1}{2} & p \\
  p & \frac{1}{2} & 1-p & \frac{1}{2}\\
  \frac{1}{2} & p & \frac{1}{2} & 1-p
  \end{pmatrix},\qquad 0<p<\frac{1}{2},
$$
and \(K^n\) denotes the \(n\times n\) adjacency matrix of a complete graph.  
This matrix \(C\) represents the connection probabilities between the four spatial segments of the small-world graphon.  
Its eigenvectors \(v_1, v_2, v_3,\) and \(v_4\) form an orthonormal basis in \(\mathbb{R}^4\):
\[
v_1=
\begin{pmatrix}
\tfrac{1}{2}\\
\tfrac{1}{2}\\
\tfrac{1}{2}\\
\tfrac{1}{2}
\end{pmatrix}, \quad
v_2=
\begin{pmatrix}
0\\
-\tfrac{1}{\sqrt{2}}\\
0\\
\tfrac{1}{\sqrt{2}}
\end{pmatrix}, \quad
v_3=
\begin{pmatrix}
-\tfrac{1}{\sqrt{2}}\\
0\\[4pt]
\tfrac{1}{\sqrt{2}}\\
0
\end{pmatrix}, \quad
v_4=
\begin{pmatrix}
-\tfrac{1}{2}\\
\tfrac{1}{2}\\
-\tfrac{1}{2}\\
\tfrac{1}{2}
\end{pmatrix},
\]
associated with the eigenvalues
\[
\lambda_1 = 2, \qquad 
\lambda_2 = \lambda_3 = 1 - 2p, \qquad 
\lambda_4 = 0.
\]
The corresponding eigenfunctions of the four-block graphon are
\[
\phi_i(x) = \sum_{j=1}^4 v_{ij}\, \mathbf{1}_{I_j}(x), 
\qquad i = 1, \dots, 4,
\]
with normalized eigenvalues \(\lambda_i / 4\).

These eigenfunctions \(\phi_1, \phi_2, \phi_3,\) and \(\phi_4\)
are identical to those obtained in Section~\ref{sec.block} for the SBM graphon, 
confirming that the small-world and stochastic block model formulations share the same spectral basis.  
Consequently, the GnSP framework developed for the SBM can be directly extended to signals defined on small-world networks.
Specifically, by projecting the network signals generated by the integrate-and-fire model simulations on small-world graphs onto the eigenfunctions derived from the four-block approximation, we performed the SIP in the same manner as in the SBM case (Fig.~\ref{Fig:figure11}E).
The simulations were performed using networks of $N=144$ neurons with the same stimulation protocol described in Section~\ref{sec:mixed}.
These results demonstrate that the proposed graphon-based framework generalizes beyond block-structured networks, providing a unified approach to signal processing on both stochastic block and small-world graphs.

\subsection{Larger block models}
\begin{figure}[ht]
%\nopagenumber
%\renewcommand{\baselinestretch}{1.0}
\hfill
\begin{center}
\includegraphics[width=0.9\textwidth]{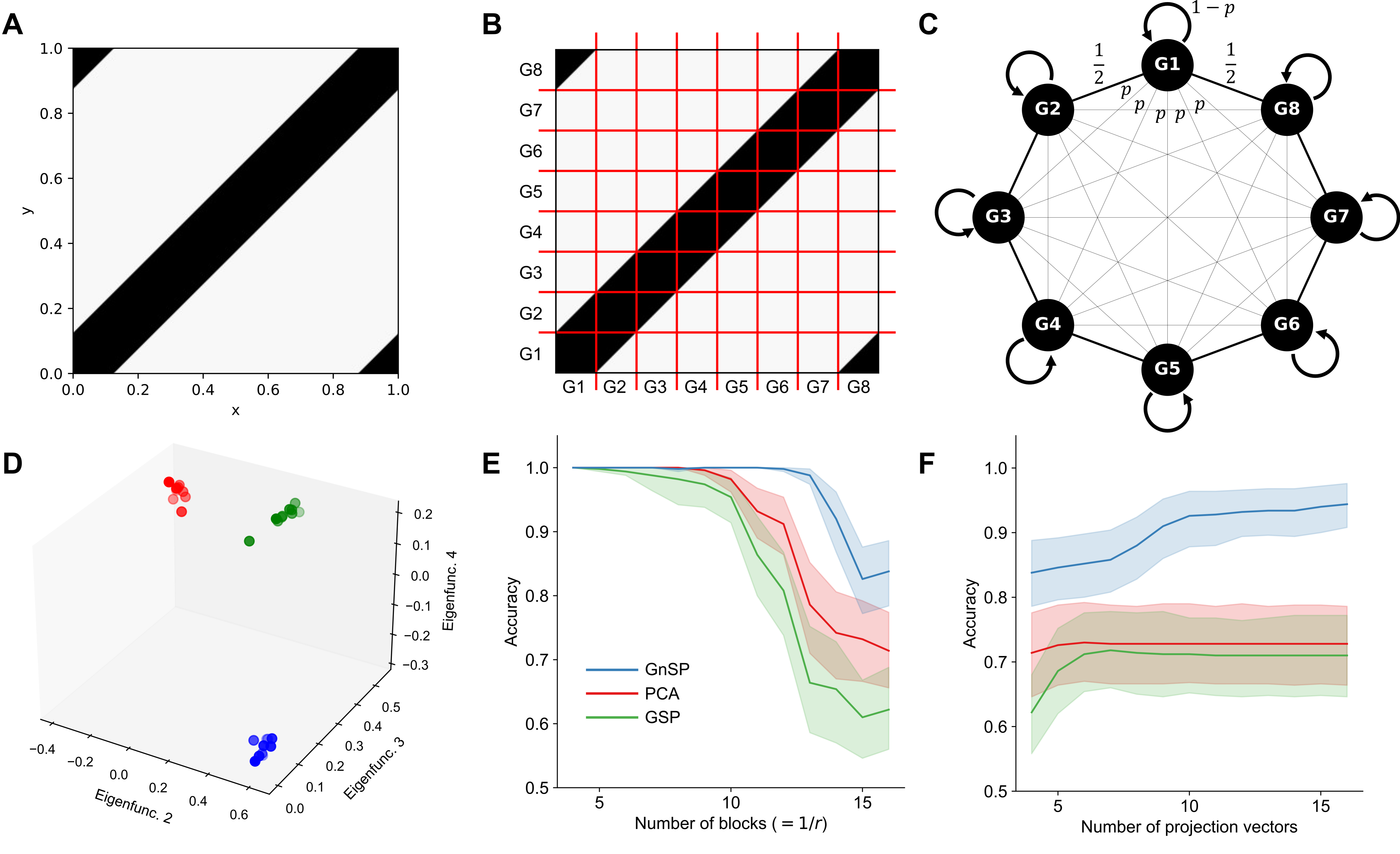}
\end{center}
\caption{\textbf{(A)} Small-world graphon with a shorter connection range of \(r = 1/8\) and \(p = 0.05\). 
\textbf{(B)} Eight-block partition of the small-world graphon. 
\textbf{(C)} Network representation of the eight-block–approximated small-world graphon. 
\textbf{(D)} GnSP embedding obtained by projecting the network activity onto the eigenfunctions 
of the block-approximated small-world graphon. 
\textbf{(E)} Classification accuracy of SIP for different block numbers (\(=1/r\)), 
based on four-dimensional projections onto the first four eigenvectors or eigenfunctions. 
\textbf{(F)} Results for the \(r = 1/16\) condition, where the number of projected eigenvectors 
was varied from 4 to 16. Shaded areas indicate 95\% confidence intervals over \(n = 50\) trials.
}
\label{Fig:figure12}
\end{figure}

To further explore the scalability of the proposed framework, we next considered the small-world graphon with a shorter connection range of \(r = 1/8\) (Fig.~\ref{Fig:figure12}A).  
By dividing the domain into eight intervals, this configuration can again be interpreted as an eight-block stochastic block model (SBM) approximation (Fig.~\ref{Fig:figure12}B,C).  
Even with an increased number of blocks, the SBM-based graphon remains stable in performing SIP under the proposed GnSP framework (Fig.~\ref{Fig:figure12}D).  
Moreover, this approach can be generalized for any \(r\), where the small-world graphon can be approximated by an SBM with approximately \(1/r\) groups.  

To quantitatively evaluate the robustness of SIP under different block resolutions, 
we employed the ridge regression procedure introduced in Section~\ref{sec.experiment} to classify the simulated signals. 
For each block configuration, the network signals were projected onto the first four eigenvectors 
(\(v_1\)--\(v_4\)), corresponding to the dominant graphon modes, 
and the resulting four-dimensional representations were used as input features for regression. 
This allowed us to assess whether the low-dimensional embeddings derived from different SBM approximations 
can still effectively separate the input stimulus conditions. 
As shown in Fig.~\ref{Fig:figure12}E, GnSP maintained higher stability and classification accuracy than both PCA and GSP, 
as the number of blocks (\(=1/r\)) increased. 
Although performance slightly decreased with larger block numbers due to the spectral cutoff issue, 
this degradation can be mitigated by increasing the projection dimensionality. 
In particular, while the four-block model used only the first four eigenfunctions, 
the sixteen-block case naturally provides sixteen eigenfunctions. 
Figure~\ref{Fig:figure12}F illustrates the results for the \(r = 1/16\) condition, 
where the number of projected eigenvectors was systematically varied from 4 to 16. 
Incorporating higher modes progressively improved classification performance, 
and GnSP consistently outperformed PCA and GSP across all tested dimensions, 
demonstrating its robustness to model granularity and network complexity.

\section{Discussion}\label{sec.discuss}
\setcounter{equation}{0}

In this study, we introduced GnSP into the analysis of neuronal population responses,
focusing on modular network architectures in both numerical simulations and real
\textit{in vitro} data. GnSP generalizes GSP by employing kernel operators derived
from graphons, measurable functions representing graphs, that are robust to changes
in network size and to variability arising from individual realizations of random graph
models \cite{Ruiz2021,Morency2021,Leus2023}. By taking advantage of these properties,
we demonstrated that graphon-based projections produce trial-invariant, separable
low-dimensional maps, effectively overcoming trial-to-trial variability and
finite-size effects, which are common limitations of standard GSP.
Numerical simulations of integrate-and-fire neural networks showed
that GnSP reliably distinguished responses across different stimulation conditions.
Furthermore, when applied to calcium imaging data from cultured modular networks,
GnSP achieved higher classification accuracy than PCA
and a previously reported reservoir-computing method \cite{Cunningham2014,Sumi2023}.

Low-dimensional representations are widely used in neuroscience to reveal structure
in high-dimensional neural activity \cite{Cunningham2014}. Techniques such as PCA and,
more recently, GSP have been applied to neural data to uncover relationships between
activity patterns and network topology \cite{Cunningham2014,Leus2023}. In particular,
GSP has been employed to identify functional modules in brain networks \cite{Atasoy2016},
filter neural signals on structural connectomes \cite{Huang2016,Glomb2020}, and detect
disease-related alterations in network organization \cite{Thomas2024,Rigoni2023}.
While these applications provide valuable insights, most focus on descriptive analysis
of neural signals on a fixed network. Standard GSP remains sensitive to the exact graph
structure, making it difficult to generalize across networks with slightly different
connectivity or size, a limitation that becomes critical in experimental settings, where variability
in network realization and measurement noise is unavoidable \cite{Shuman2013,Ortega2018}.

The SIP, the inverse problem of inferring the stimulus pattern $s$ from observed neural responses $f$, is of fundamental importance in neuroscience and
closely relates to neural decoding, where one seeks to reconstruct sensory
inputs or motor outputs from neural activity \cite{Glaser2020,Cunningham2014}.
Many decoding approaches do not take advantage of the underlying network topology,
focusing instead on direct statistical mappings between activity and stimuli \cite{Glaser2020}.
By incorporating network organization into the decoding process, our framework can
address problems in which the spatial arrangement of connectivity shapes neural responses,
a scenario that is very common in neural systems.

The present work represents the first application of GnSP to experimental neuroscience.
By providing size- and topology-robust coordinates, our framework enables stable
low-dimensional representations even when the underlying network structure varies across
realizations \cite{Ruiz2021,Morency2021}. Although the improvement in classification accuracy
over PCA did not reach statistical significance in the available dataset, the observed effect
size suggests that the graphon-based method has the potential to enhance discriminability
in larger-scale studies. The relatively small sample size ($n=21$) in the present analysis limits
statistical power, highlighting the importance of validating these findings on more
extensive datasets. It is also worth noting that the present approach is much simpler
to implement compared to PCA or reservoir computing techniques.
Furthermore, our numerical simulations, repeated across different values of network size 
$n$ and edge density $\alpha$, provide additional evidence of the robustness and
efficiency of GnSP for the SIP.

We finally demonstrated that the GnSP framework can be extended beyond the four-block model to more complex network structures, including small-world and higher-block graphons. By showing that small-world connectivity can be approximated by a block representation sharing the same spectral basis, and that the framework remains robust across increasing model granularity, we established the scalability of GnSP with respect to network topology and resolution. These results confirm that the proposed method is not restricted to simple modular networks but provides a general and adaptable foundation for graph-based signal analysis.

Looking ahead, GnSP offers a systematic framework for the analysis of signals defined
on networks. It is particularly well suited to engineered \textit{in vitro} preparations
\cite{Albers2016,Habibey2024,WinterHjelm2023,Sumi2023,Murota2025,Monma2025},  where microfabrication and patterning
allow precise control of mesoscale structure and repeated instantiation of
comparable network architectures. In this setting, graphon-based operators
enable size- and noise-robust analyses for decoding, neural signal propagation,
and perturbation studies. For \textit{in vivo} applications, as anatomical
brain networks become denser and more consistent across scales,
graphon spectra and filters provide an efficient means of comparing
signals across subjects and regions
while accounting for differences in network size and connectivity
\cite{Huang2016,Glomb2020}.

In view of its potential applications, our work suggests that GnSP is
a promising analytical framework for biological network data,
bridging the gap between theoretical developments in network
signal processing and real-world neuroscience. Broader applications
and larger datasets will be essential to fully establish its advantages.
Future studies may also explore integrating GnSP with other machine
learning or
graph-based techniques to further advance neural data analysis.

\subsection*{Acknowledgments}
The work of TS was supported by the JST Moonshot R\&D Program (JPMJMS2023-41) and partially by MEXT Grant-in-Aid for Transformative Research Areas (A) “Multicellular Neurobiocomputing”
(24H02332).
The work of GSM was partially supported by NSF DMS Award \#2406941. GSM is grateful to
  Hayato Chiba for the invitation to visit the Advanced Institute for Materials Research at
  Tohoku University, and to all members of Chiba’s Lab for their hospitality.

% \bibliographystyle{unsrt}
% \bibliography{gsp}     

\end{document}